# Spatial distributions of Balmer series emissions in divertor plasmas with tungsten targets in different magnetic configurations on the EAST superconducting tokamak

**Kunpei Nojiri (野尻訓平)[1], Naoko Ashikawa (芦川直子)[2,3], Yasuhiro Suzuki (鈴木康浩)[2,3], Yaowei Yu (余耀伟)[4], Zhenhua Hu (胡振华)[4], Fang Ding (丁芳)[4], Liang Wang (王亮)[4], Lingyi Meng (孟令义)[4], Xiahua Chen (陈夏华)[4], Jie Huang (黄杰)[2], Zhongshi Yang (杨钟时)[4], Tetsutaro Oishi (大石鉄太郎)[2,3], Mizuki Sakamoto (坂本瑞樹)[1], Jiansheng Hu (胡建生)[4], Guangnan Luo (罗广南)[4]**

[1]Plasma Research Center, University of Tsukuba, 1-1-1 Tennodai, Tsukuba 305-8577, Japan
[2]National Institute for Fusion Science, National Institutes of Natural Sciences, Oroshi-cho, 322-6, Toki 509-5292, Japan
[3]Department of Fusion Science, Graduate University for Advanced Studies, SOKENDAI, Oroshi-cho, 322-6, Toki 509-5292, Japan
[4]Institute of Plasma Physics, Chinese Academy of Sciences, Hefei 230031, People's Republic of China

E-mail: nojiri_kunpei@prc.tsukuba.ac.jp

## Abstract

The characteristics of wall recycling with different divertor configurations were investigated in this study, focusing on the observations of the spatial distributions of deuterium atomic emissions in the Balmer series ($D_\alpha$, $D_\beta$, $D_\gamma$, and $D_\delta$) with different magnetic field configurations in the Experimental Advanced Superconducting Tokamak. The observed $D_\alpha$ and $D_\beta$ emissions were primarily relatively close to the divertor targets, while the $D_\gamma$ and $D_\delta$ emissions were primarily relatively close to the X-point. The distributions of the emissions close to the divertor targets and X-point changed differently depending on the divertor configuration. These experimental results indicate that the linear comparison of parameters based on an assumption of similarity of profile shapes in different configurations is insufficient for understanding particle recycling in divertor plasmas. This is because the shape of the density profile of the recycled deuterium atoms and/or the electron density and temperature may change when the magnetic configuration is altered.

## 1. Introduction

Hydrogen isotope recycling at divertor targets plays important roles in core plasma performance [1–3]. Injected fuel particles (hydrogen isotopes) are ionized, and some of the fuel ions eventually reach walls. After reaching the walls, the ions are reflected as neutral atoms or absorbed by the walls. Some of the absorbed ions are re-emitted in the form of neutral molecules. These recycled neutral atoms and molecules may re-enter the plasma and cause interactions such as refueling [4]. Recycling control will be even more important in future fusion devices with long-pulse operation, such as ITER [5, 6] and DEMO [7, 8], to maintain core plasma performance. The material species of the walls is an important factor in determining the recycling characteristics [9]. ITER will use tungsten (W) as the divertor material, which is also a candidate material for DEMO. The recycling characteristics additionally depend on wall conditions such as retention, temperature, and deposition [10]; the divertor geometry and configuration [2, 11]; and plasma parameters. Since these effects on recycling change with the position in the divertor region, it is necessary to investigate the parameters related to recycling at various positions. Thus, it is important to assess the spatial effects of the divertor configuration on the recycling characteristics with W divertor plates.

For recycling studies in tokamaks, hydrogen isotope atomic emissions, especially Balmer series emissions such as $D_\alpha$ and $D_\beta$, are widely used. Some of the neutral atoms generated by reflection at the walls and dissociation of the re-emitted molecules from the walls are excited by electrons, and Balmer series emissions are observed. Thus, atomic emissions are effective monitors of recycling neutrals. These observed emissions are often the results of integration along the lines of sight (LOSs) of spectrometer systems, meaning that there is uncertainty in the locations of the recycled atoms. To evaluate local emissions, techniques such as tomography [12] and Zeeman patterns in spectral shapes [13] are used. Meanwhile, for emissions from higher excited states such as $D_\gamma$ and $D_\delta$, their ratios to $D_\alpha$ are widely utilized as indicators of volumetric recombination [14]. In contrast, in ionizing plasma, the ratio becomes larger with decreasing electron density $n_e$ [15], while its dependence on the electron temperature $T_e$ is relatively small. In the scrape-off layer (SOL)/divertor region, since $n_e$ changes significantly with position, the spatial distributions of the different Balmer series emissions in ionizing plasma can be compared to estimate the locations of emissions from recycled atoms. This method is simple and involves relatively basic spectroscopic system setups. Therefore, observations of the spatial distributions of Balmer series emissions with different configurations are effective for elucidating the spatial effects of the divertor configuration on the recycling characteristics.

In the Experimental Advanced Superconducting Tokamak (EAST), a water-cooled W monoblock is used for the upper divertor [16]. Recently, utilizing superconducting coils, heating systems and feedback control systems, H-mode operation over 100 s has been achieved with an upper single-null (USN) configuration [16–18]. Meanwhile, it has been observed that the strike point (SP) positions at the W divertor (configuration) alter the exhaust efficiency of the recycled neutrals, which is a spatially averaged characteristic of recycling, focusing on the density decay of the bulk plasma after gas puffing is stopped [18, 19]. To elucidate the spatial effects of the divertor configuration on the recycling characteristics with a W divertor, the spatial distributions of Balmer series emissions in the divertor region with different configurations were investigated in this study.

## 2. Experimental setup

Deuterium plasma discharge is conducted with a USN divertor configuration in EAST. The upper divertor is equipped with water-cooled W/Cu plasma-facing components [16]. An in-vessel cryopump is installed for particle exhaust through a vacuum pumping slot at the corner of the upper outer (UO) divertor region, as shown in figure 1(a). The exhaust pressure is measured by a cold cathode ionization gauge.

Multichannel spectroscopic systems of the deuterium Balmer series ($D_\alpha$, $D_\beta$, $D_\gamma$, and $D_\delta$) are installed. Figure 1(a) shows the LOSs mapped to a poloidal cross-section. The LOSs of the system start from an optical vacuum window located below the equatorial plane and extend to UO divertor targets at different toroidal positions. Three sets of optical fiber systems are used: $D_\alpha$, $D_\beta$, and $D_\gamma$ and $D_\delta$. Hence, different LOSs are shown in figure 1(b). The obtained counts in the spectroscopy systems are integrated values along the relatively complicated LOSs. However, it is considered that emissions from neutral deuterium atoms are mainly produced between the X-point and UO divertor targets. The system for $D_\alpha$ measurement included a photomultiplier tube; the $D_\gamma$ and $D_\delta$ emissions were measured by a spectrometer with an electron-multiplying charge coupled device (EMCCD) [20]; and an additional multichannel spectrometer system with an EMCCD was used for the $D_\beta$ measurements. Relative intensity calibration of the $D_\gamma$ and $D_\delta$ measurements was performed previously to obtain their spatial distributions. For the $D_\alpha$ and $D_\beta$ measurements, precise relative calibration has not been conducted, but these signal intensities were adjusted to reduce differences in sensitivity among the observation channels. On the UO divertor targets, 13 Langmuir probes (LPs) are installed in one poloidal section [21], as shown in figure 1(b), and those at port O were used in this study. Data acquisition of the LPs and $D_\alpha$ was conducted about every 0.02 ms and 0.01 ms, respectively. The exposure times for the $D_\beta$ measurements and the $D_\gamma$ and $D_\delta$ measurements were 5 ms and 10 ms, respectively, and sampling rates were about 100 Hz.

## 3. Experimental results and discussion

The plasma experiment was performed with the USN configuration, a toroidal magnetic field $B_T$ of ~2.4 T (normal field), a plasma current $I_p$ of ~400 kA, and lower hybrid wave heating with a power of ~1 MW in L-mode (shot # 93922).

For plasma operation with different divertor magnetic configurations under similar background conditions of plasma-facing walls, the SP was swept at the UO divertor. The line-averaged density $\bar{n}_e$ was set to ~3 × 10$^{19}$ m$^{-3}$ by a density feedback control via supersonic molecular beam injection. As shown in figure 2, $I_p$ and $\bar{n}_e$ were kept almost constant from time $t$ = 3.3 s to 6.3 s, and the SP position calculated by EFIT was shifted from a distance from the UO divertor corner $d$ of 8 cm to 12 cm along the divertor targets. Thus, the divertor SP moved away from the vacuum pumping slot.

From $t \approx 3.4$ s to 5.0 s, the divertor neutral pressure $P_{n,div}$ decreased, which is consistent with the experimental results on particle exhaust efficiency optimization regarding the SP position at the divertor targets [18, 19]. However, from $t \approx 5.0$ s to 6.5 s, although the SP position continued moving away from the slot, $P_{n,div}$ increased. This finding is different from the results published in the abovementioned literature, where the SP was located at $d = 5$–8 cm. Based on this difference, it was assumed that the neutral pressures do not simply depend on the SP position from the vacuum pumping slot. Hence, additional investigations of other effects on $P_{n,div}$ are required.

In this study, the SP position was confirmed by experimental data obtained by LPs on divertor targets in addition to numerical calculations performed using EFIT. Figure 3 shows the separatrices calculated by EFIT at $t = 3.3$, 3.6, 4.8, 5.4, and 6.3 s at plasma discharge #93922. At $t = 3.3$, 3.6, 4.8, 5.4, and 6.3 s, $d = 8$, 9, 10, 11, and 12 cm, respectively. When the SP position shifts from the inner to outer side, the position of the X-point also moves radially outward.

Figure 4(a) shows the time evolution of the spatial distribution of the ion saturation current $I_s$ measured by LPs on divertor targets. Figures 4(b) and (c) present the spectroscopy data of $D_\gamma$ and $D_\delta$ as spatial distributions projected onto divertor targets. $I_s$ in figure 4(a) was obtained using values averaged over every 10 ms, and the intensities of $D_\gamma$ and $D_\delta$ acquired in one exposure time (10 ms) are plotted in figures 4(b) and (c), respectively. This time is sufficiently short to follow the change in SP position. When the SP moves from $d = 8$ cm to 12 cm, the peak positions of $I_s$, $D_\gamma$, and $D_\delta$ also move outward. Some of the peak positions of $I_s$ almost exactly match the SP positions. On the other hand, the positions corresponding to higher intensities of $D_\gamma$ and $D_\delta$ occur more than 2 cm outside the SP according to the projected positions on the divertor targets. This LOS with the peak intensities of $D_\gamma$ and $D_\delta$ passes above the X-point, as shown in figure 1(b).

Figure 5 presents the time evolution of the intensities of $D_\alpha$ and $D_\beta$ measured by multichannel spectroscopy. The positions, as shown from 6.1 cm to 20.7 cm, are the projected positions of the integrated signals along the LOSs to the divertor targets, as depicted in figure 1(b). The intensities of $D_\alpha$ averaged over every 2 ms and the intensities of $D_\beta$ corresponding to every 5 ms exposure are plotted. When the SP position is around 8 cm, higher intensities of $D_\alpha$ and $D_\beta$ are observable at the projected position of 10 cm. This tendency is similar to that of the peaks of $I_s$ around the SP position of 8 cm in figure 4(a). After that, with outward movement of the SP, the intensities of $D_\alpha$ and $D_\beta$ at the projected positions of 10 cm and 12 cm decrease. On the other hand, when the SP moves from 10 cm to 12 cm, the intensities at the projected positions of 15–21 cm increase. Thus, with an SP position of 12 cm, the projected positions of higher intensities of $D_\alpha$ and $D_\beta$ seem to be located in the range of 15–21 cm. These findings also suggest that the projected position of higher intensity moves 5–11 cm outward when the SP moves from $d = 8$ cm to 12 cm.

Figure 6(a) shows the spatial distributions of $I_s$ on the divertor targets. Figures 6(b) and (c) present the projected spatial distributions of $D_\gamma$ and $D_\delta$ with SP positions of $d = 8$, 9, 10, 11, and 12 cm ($t = 3.3$, 3.6, 4.8, 5.4, and 6.3 s). Figure 6 plots the values averaged over every 100 ms and provides the details of the spatial distributions. When $d = 8$, 10, and 12 cm, the peak positions of $I_s$ are shifted up to 1 cm from the SP position, including consideration of the spatial resolution error due to the limited number of LPs. Further, $I_s$ peaks are not observable near the SP position when $d = 9$ and 11 cm. Judging from the SP position, the peak positions of $I_s$ are considered to be in the range from 7 cm to 14 cm, and $I_s$ positions greater than 14 cm are not considered to be peaks.

In the projected spatial distributions of $D_\gamma$ and $D_\delta$ on the divertor targets, peak positions occur at 13–17 cm when the SP position is 8–12 cm, and the projected peak positions are about 4 cm outward from the SP position. The projected spatial distribution of $D_\gamma$ shows a widely expanded tail on the inward side when the SP is at $d = 8$ cm, but the $D_\gamma$ distributions when the SP is at $d = 12$ cm do not have such larger tails below 12 cm. Therefore, the $D_\gamma$ intensities for the two SPs, which correspond to different magnetic configurations, suggest different spatial profile shapes. The difference between the positions of the maximum intensities on the divertor targets in the cases in which the SP is located at $d = 8$ and 12 cm is about 2 cm. The signal-to-noise ratio of the $D_\delta$ intensity is low due to the low count of $D_\delta$, whose maximum count is only about 30; thus, the errors for the peak positions are greater than those for the $D_\gamma$ peaks. However, the peak position of $D_\delta$ in the case in which the SP is located at $d = 8$ cm is at least inside the peak position corresponding to $d = 12$ cm, and the peak positions differ by up to 4 cm. Note that the spatial profile shapes of $D_\gamma$ and $D_\delta$ are significantly different from that of $I_s$. $I_s$ is related to the local product of $n_e$ and the ion sound speed at the divertor plate. On the other hand, the intensities of $D_\gamma$ and $D_\delta$ are related to the integrated values of the local product of $n_e$, the deuterium atomic density, and the rate coefficients of the excitation reactions along the entire LOS, including the divertor plates and X-point. The analyses of the profile shape differences with these parameters taken into account are planned.

As discussed in relation to figure 5, the projected positions of higher intensities of $D_\alpha$ and $D_\beta$ are around 10 cm and in the range of 15–21 cm when the SP is located at $d = 8$ and 12 cm, respectively. The peak positions of the projected $D_\gamma$ and $D_\delta$ intensities in figures 6(b) and (c) are around 13 and 17 cm when the SP is located at $d = 8$ and 12 cm, respectively. All the projected positions of the deuterium atomic emissions are outside the SP positions. Following the LOSs in figure 1(b), it can be understood that as the projected position becomes more distant, the LOS crosses a point further upstream in the separatrix. If the emissions of deuterium atoms are released along divertor magnetic field lines from the X-point to the divertor targets, the emissions between the X-point and the vicinity of the divertor targets are dominant, rather than those in the vicinity of the divertor targets. In addition, when the SP is located at $d = 8$ cm, the projected positions of the $D_\gamma$ and $D_\delta$ peaks are farther away than those of higher $D_\alpha$ and $D_\beta$ intensities. This finding suggests that $D_\gamma$ and $D_\delta$ emissions are dominant on the X-point side of the above region, while $D_\alpha$ and $D_\beta$ emissions are dominant on the divertor side. In ionizing plasma, intensity ratios such as $D_\gamma/D_\beta$ and $D_\delta/D_\beta$ considerably decrease with $n_e$ and moderately increase with $T_e$ in the $n_e$ range of $10^{16}$–$10^{20}$ m$^{-3}$ [15]. In the SOL/divertor region of EAST, $n_e$ has generally been observed to be within this range [21, 22]. In this experiment, it is considered that $n_e$ decreased and $T_e$ increased from the divertor targets to the X-point. Therefore, it can be considered that higher $D_\alpha$ and $D_\beta$ emissions occurred relatively close to the divertor targets, while those of $D_\gamma$ and $D_\delta$ occurred relatively close to the X-point. It should be noted that the projected positions of higher $D_\alpha$ and $D_\beta$ intensities moved by at least 5 cm and those of $D_\gamma$ and $D_\delta$ moved by up to 4 cm. This difference indicates that the distributions of the strong emissions near the divertor targets and those near the X-point change differently depending on the magnetic field configuration. It can be also understood that the change in the profile shape of $D_\gamma$ means that the profile shape of the emission near the X-point changes with the configuration. These results suggest that the distribution of the density of recycled deuterium atoms and/or $n_e$ and $T_e$ changed, since the emission intensities are functions of the atomic density $n_e$ and a rate coefficient of excitation that depends on $T_e$. Therefore, to understand the spatial distribution of the density of recycled atoms, it is necessary to evaluate local emissions using experimental data, i.e., three-dimensional LOS and magnetic field data, and numerical simulation data for the edge plasma.

In EAST, wall recycling has been investigated in relation to the vacuum pumping efficiency, effects of Li coating, and plasma-facing materials. In the present study, the spatial distributions of deuterium atomic emissions in the edge plasma regions were observed, and the characteristics of the distributions corresponding to different magnetic configurations were identified. The spatial profile shapes of the density of recycled atoms and/or $n_e$ and $T_e$ could change when the magnetic configuration changed. Consequently, linear comparison of parameters with different configurations assuming similarity of profile shapes is insufficient. Future devices such as the China Fusion Engineering Test Reactor [23] and ITER have large plasma-facing areas and their operations are planned to be long-pulse, and wall recycling control remains an important issue for density control. Therefore, it is important to clarify the spatial behaviors of the local emissions related to wall recycling during plasma experiments to achieve density control during steady-state operations, as shown in this study.

## 4. Conclusions

This paper focused on the spatial effects of the divertor configuration on the recycling characteristics in the W divertor region. So far in EAST, it has been confirmed that the configuration changes the exhaust efficiency of the recycled neutrals throughout the W divertor region, which is a spatially averaged characteristic. In this study, the spatial distributions of deuterium atomic emissions in edge plasma regions were observed, and the characteristics of their spatial distributions with different magnetic configurations were identified. Based on the observations, the differences among the $D_\alpha$, $D_\beta$, $D_\gamma$, and $D_\delta$ distributions were determined, and it can be considered that the distributions of the density of recycled deuterium atoms and/or $n_e$ and $T_e$ change with the magnetic configuration. In particular, $D_\alpha$ and $D_\beta$ emissions are dominant relatively close to the divertor targets, while $D_\gamma$ and $D_\delta$ emissions are dominant relatively close to the X-point region rather than close to the divertor targets. The distributions of the emissions from deuterium atoms relatively close to the divertor targets and X-point change differently with the magnetic field configuration. The shape of the emission profile near the X-point also changes clearly. These experimental results indicate that linear comparison of parameters assuming similarity of profile shapes with different configurations is insufficient, because the spatial profile shapes of the density of recycled atoms and/or $n_e$ and $T_e$ may change depending on the configuration.

In the future, it is planned to compare the experimental results with numerical simulation results for edge plasma parameters using three-dimensional LOS and magnetic field data to evaluate local deuterium atom emissions and to improve understanding of the spatial distribution of the density of recycled deuterium atoms.

## Acknowledgments

This work is supported by JSPS-CAS Bilateral Joint Research Projects,"Control of wall recycling on metallic plasma facing materials in fusion reactor," 2019-2022, (GJHZ201984 and JPJSBP120197202).

**Figures**

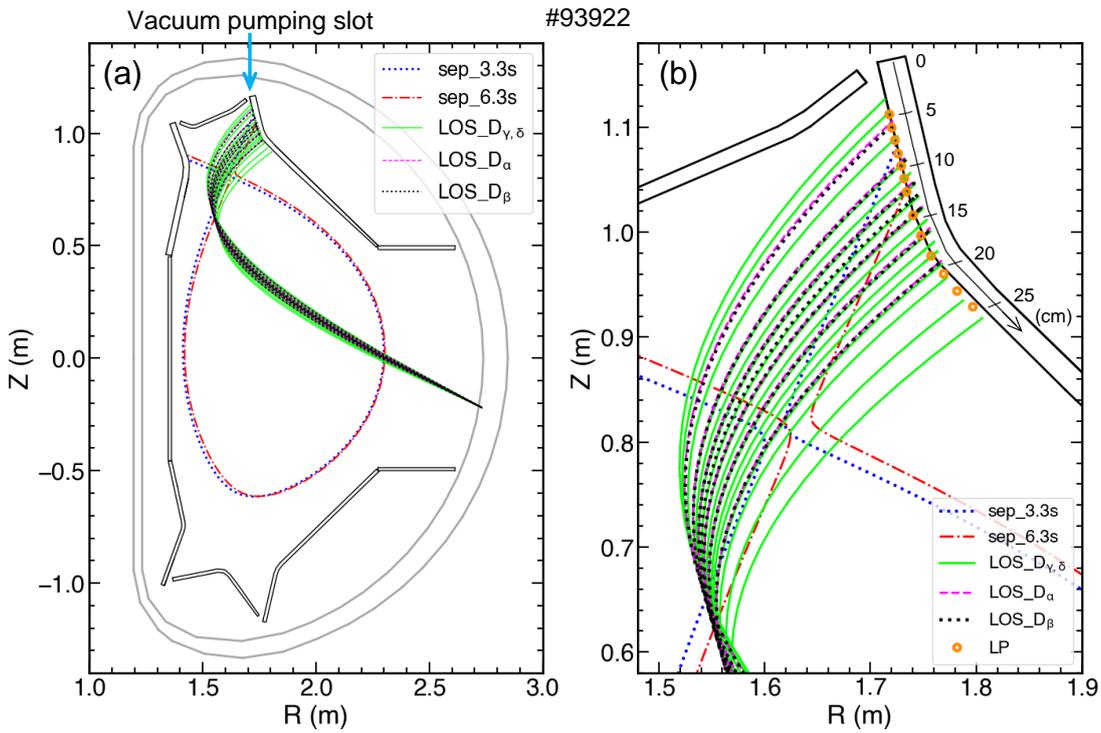

**Figure 1.** (a) Poloidal cross-section of EAST with separatrices and LOSs mapped onto the poloidal plane, which are adopted by the multichannel visible spectroscopy systems for deuterium Balmer series measurements in the upper W divertor. (b) Enlarged view of the divertor region with the positions of the LPs on the divertor targets. Both graphs show two kinds of separatrices obtained by EFIT at different times, $t$ = 3.3 and 6.3 s, in #93922.

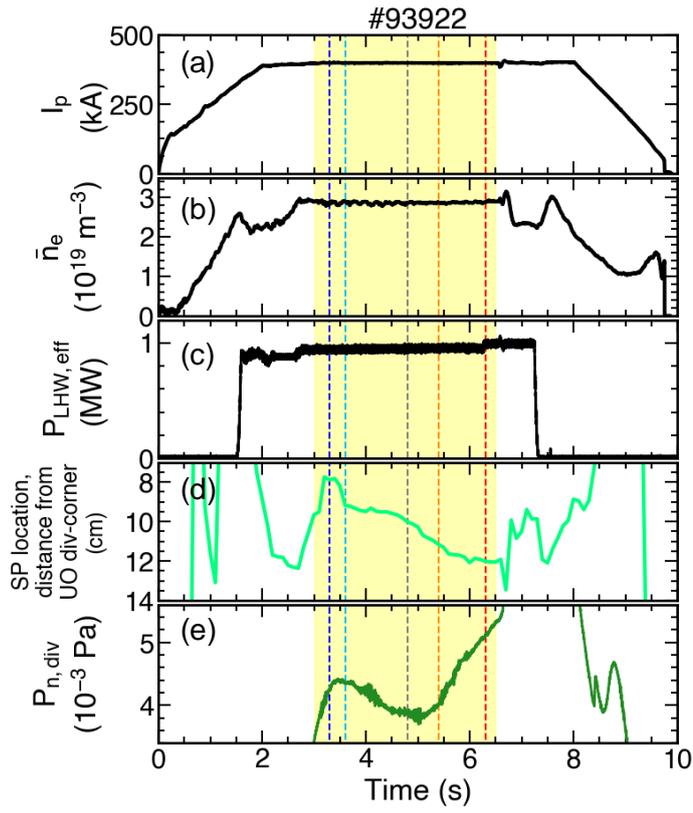

**Figure 2.** Time evolution of (a) $I_p$, (b) $\bar{n}_e$, (c) effective (injected - reflected) power of lower hybrid wave, (d) SP position as described by $d$, and (e) $P_{n,div}$.

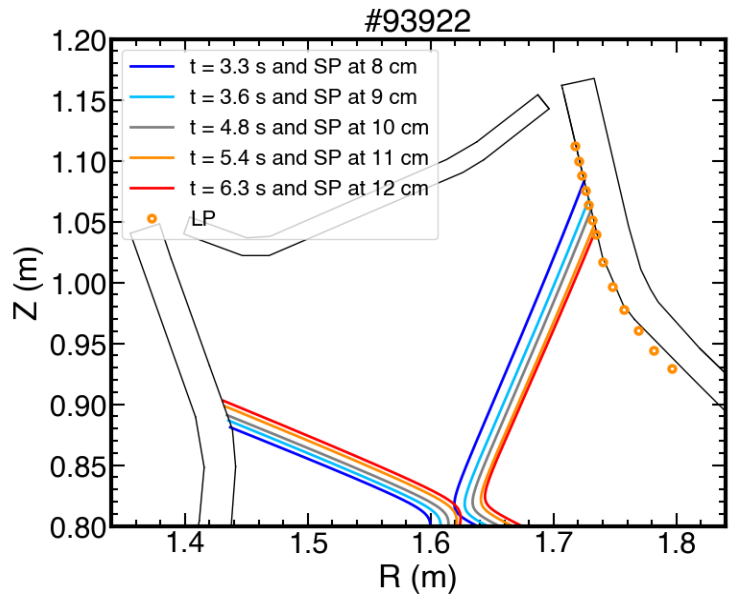

**Figure 3.** Separatrices around the divertor region at each selected time in #93922 and positions of LPs on the UO divertor with the SP.

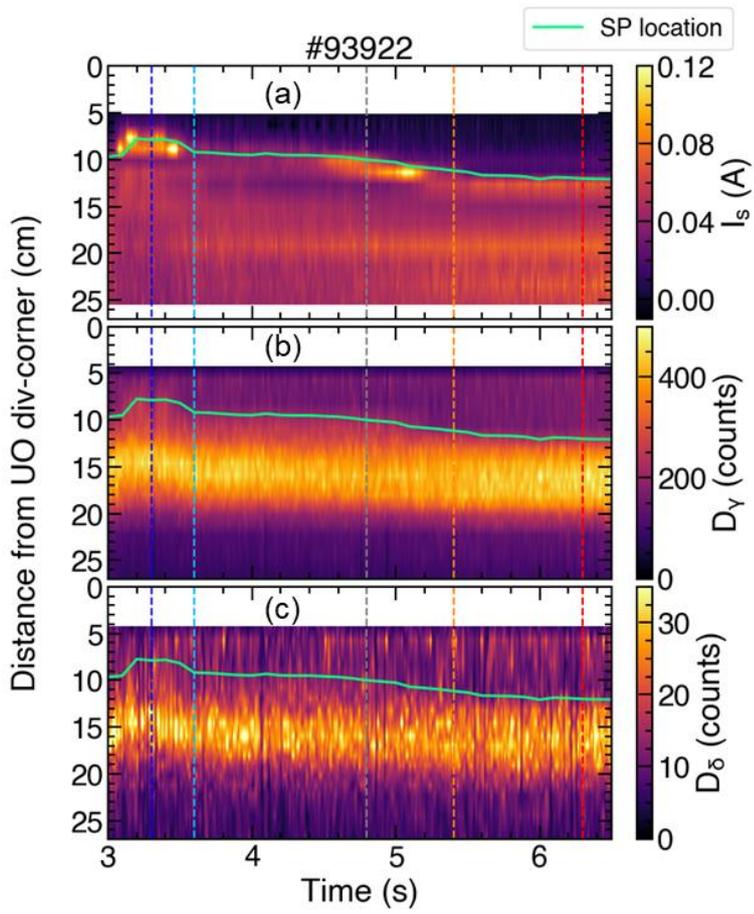

**Figure 4.** (a) Time evolution of the spatial distribution of $I_s$ on the UO divertor. Projected spatial distributions of (b) $D_\gamma$ and (c) $D_\delta$ intensities on the divertor targets. The green line shows the position of the divertor SP on the UO divertor as described by $d$.

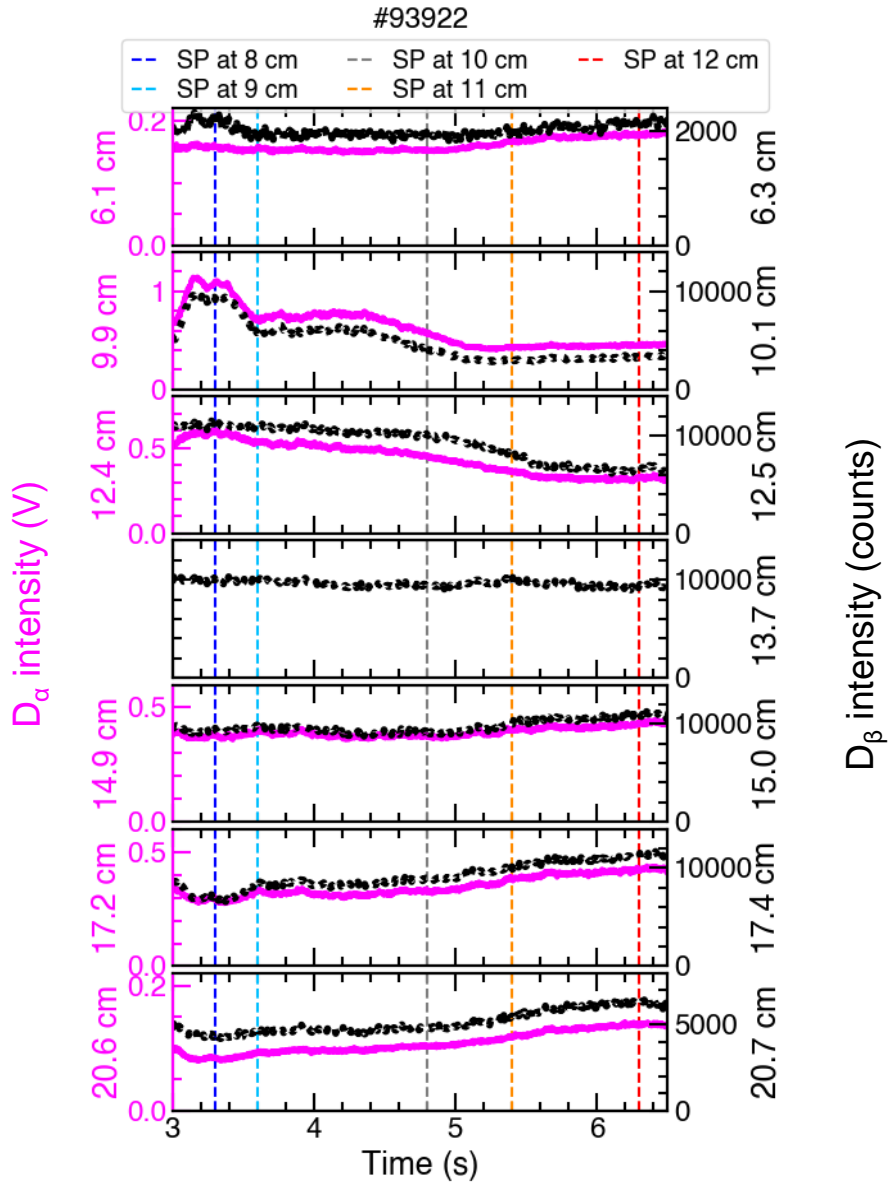

**Figure 5.** Time evolution of $D_\alpha$ (left) and $D_\beta$ (right) intensities measured by multichannel spectroscopy. Each position, as shown from 6.1 cm to 21 cm, is the projected position of an LOS to the divertor targets.

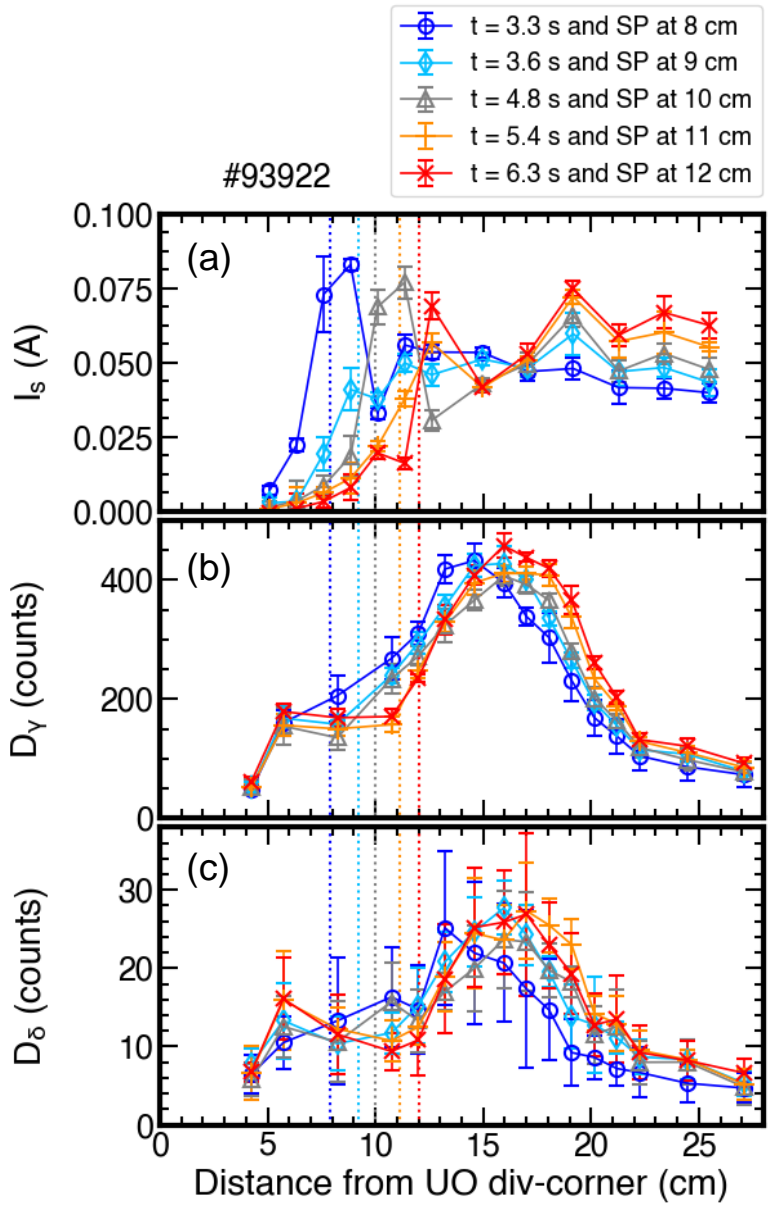

**Figure 6.** (a) Spatial distributions of $I_s$ on divertor targets. Projected spatial distributions of (b) $D_\gamma$, and (c) $D_\delta$ intensities on the UO divertor at each selected time. The dashed lines show the positions of the divertor SP on the UO divertor as described by $d$.